\documentclass{segabs}

\begin{document}

\title{Extremely Weak Supervision Inversion of Multi-physical Properties}

\renewcommand{\thefootnote}{\fnsymbol{footnote}} 

\author{Shihang Feng$^1$, Peng Jin${^1}$, Xitong Zhang$^{1}$, Yinpeng Chen$^{2}$, David Alumbaugh$^{3}$, Michael Commer$^{3}$ and Youzuo Lin$^1$
	\\
$^{1}${Los Alamos National Laboratory, USA}\\
$^{2}${Microsoft Research, USA}\\
$^{3}${Lawrence Berkeley National Laboratory, USA}\\
}

\footer{Example}
\lefthead{Feng et al.}
\righthead{WS-MGI}

\maketitle

\begin{abstract}
Multi-physical inversion plays a critical role in geophysics. It has been widely used to infer various physical properties~(such as velocity and conductivity). Among those inversion problems, some are explicitly governed by partial differential equations~(PDEs), while others are not. Without explicit governing equations, conventional multi-physical inversion techniques will not be feasible and data-driven inversion requires expensive full labels. To overcome this issue, we develop a new data-driven multi-physics inversion technique with extremely weak supervision. Our key finding is that the pseudo labels can be constructed by learning the local relationship among geophysical properties at very sparse well-logging locations. We explore a multi-physics inversion problem from two distinct measurements~(seismic and EM data) to three geophysical properties~(velocity, conductivity, and CO$_2$ saturation). Our results show that we are able to invert for properties without explicit governing equations. Moreover, the label data on three geophysical properties can be significantly reduced by 50 times~(from 100 down to only 2 locations).

\end{abstract}

\section{Introduction}
Geophysical and fluid properties (such as velocity, conductivity and $\mathrm{CO_{2}}$ saturation) provide structural and numerical information for various geophysical applications, e.g. assessment of oil and gas reservoirs and sequestration of $\mathrm{CO_{2}}$~\citep{lucia2003carbonate}. These properties are obtained from surface-based geophysical measurements including seismic~\citep{yilmaz2001seismic}, electromagnetics (EM)~\citep{zhdanov2009geophysical}, gravity~\citep{li19983}, etc, by geophysical inversion. 

These inversion problems have been studied \textit{separately} along two directions: physics-driven and data-driven. The physics-driven methods~\citep{zhdanov2000electromagnetic,virieux2009overview,feng2017skeletonized,feng2019transmission+,chen2020multiscale} are applicable for seismic$\rightarrow$velocity and EM$\rightarrow$conductivity by leveraging the known PDE, which is converted as a forward modeling operator such that the input, velocity or conductivity properties, is a function of seismic or EM data as output. Based on the forward modeling, velocity and conductivity can be iteratively optimized. The data-driven methods apply to the inversion problems by leveraging deep neural networks to learn a correspondence from geophysical measurements to geophysical properties~\citep{araya2018deep,wu2019inversionnet,jin2020cyclefcn,feng2021multiscale_style}. This type of work requires a large amount of paired geophysical measurements and geophysical properties to train the network. 

\begin{figure}[ht]
\centering
\includegraphics[width=0.9\columnwidth]{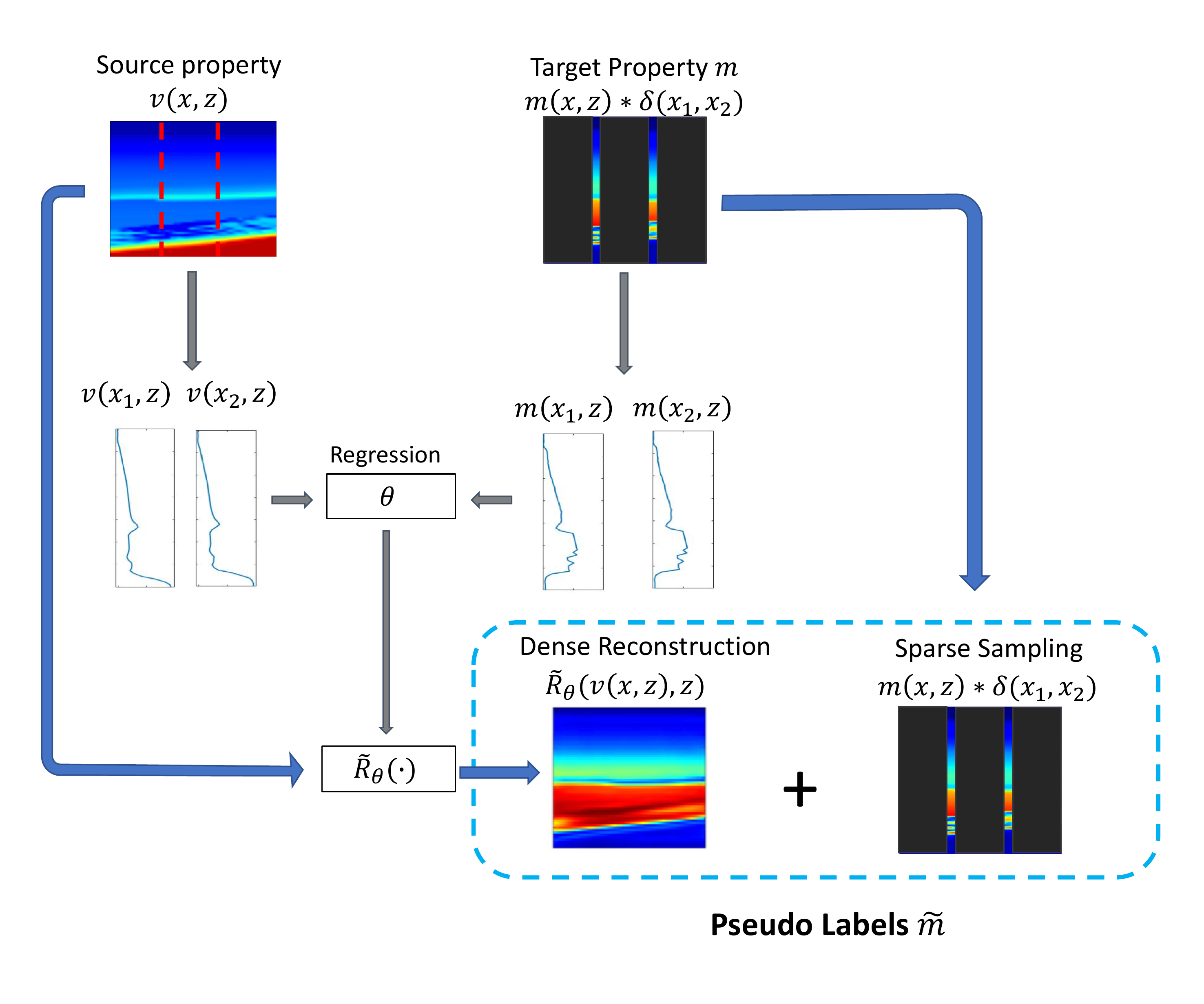}
\caption{Schematic illustration of our proposed method, which generate the pseudo labels $\tilde{m}$ from the sparse samplings of the target property $m$ and the full labeling of the source property $v$.}
\label{fig:WS-MGI.pdf}
\end{figure}


Properties without explicit governing equations can be obtained with supervised data-driven methods. However, the acquisition of the labeled data is extremely expensive, only sparse labeled data can be acquired in the field experiments.~\cite{sun2020deep} firstly present a joint inversion that reconstructs salt geometry by combining seismic and electromagnetic data. However, this method still relies on a large amount of labeled data, which is impossible to be obtained in the real case.

In this work,  we shift the data-driven inversion paradigm to \textit{jointly} address the following three inversion problems with \textit{extremely weak} supervision. The three inversion problems are as follows: (a) seismic$\rightarrow$velocity to recover velocity models from seismic data, (b) EM$\rightarrow$conductivity to recover conductivity models from EM data, and (c) seismic/EM$\rightarrow$CO\textsubscript{2} to recover CO\textsubscript{2} saturation models from seismic and EM data. In the first step, a single-physics inversion is performed in an unsupervised way. In the second step, we construct the pseudo labels by approximating the relationship between the geophysical properties, which enable the inversion of the properties that do not have explicit governing equations. The requirement of the multi-physics labeled data is greatly reduced. We name our multi-physics method as Weakly Supervised Multiple Geophysics Inversion (WS-MGI) and evaluate our methodology on the Kimberlina reservoir data~\citep{Development-2021-Alumbaugh}. These numerical results demonstrate that WS-MGI can accurately reconstruct the subsurface structures with sparsely labeled data.

\section{Theory}

\begin{figure*}[ht]
\centering
\includegraphics[width=1.85\columnwidth]{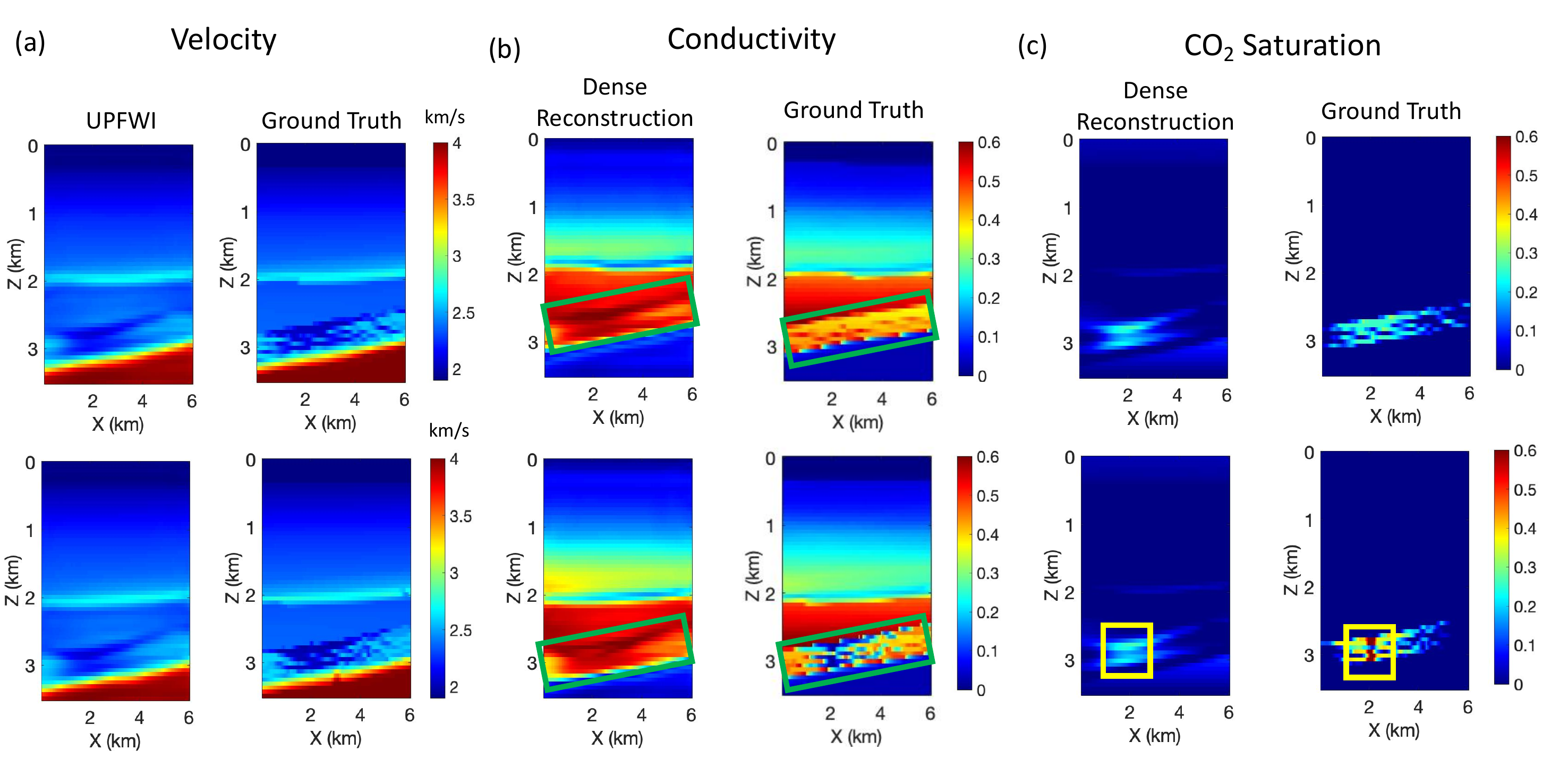}
\caption{(a) Velocity models given by UPFWI. (b) Conductivity dense reconstruction and true conductivity model. (c) $\mathrm{CO_2}$ saturation dense reconstruction and true $\mathrm{CO_2}$ saturation model.}
\label{fig:SVR_data.pdf}
\end{figure*}

\subsection{ Weakly Supervised Multiple Geophysics Inversion}

Here we proposed a Weakly Supervised Multiple Geophysics Inversion (WS-MGI) method to invert multi-physics properties (source property $v$ and target property $m$) with sparse samplings. The samplings are well logs $m(x=x_k,z)$, where $x_k$ is the drilling location. The source property $v$ is a property related to the geophysical measurements with PDEs while the target property $m$ is the property without explicit governing equations. WS-MGI is implemented in two stages.



\textbf{Stage 1. Unsupervised Single Geophysical Inversion:} Stage 1 is an unsupervised inversion for the single geophysical property $v$, which had already been proposed by~\cite{jin2021unsupervised} as Unsupervised Physical-Informed Full Waveform Inversion (UPFWI) for velocity models as $v$. 

\textbf{Stage 2. Pseudo Labels Building and Training:}
 To build the pseudo labels in Stage 2, we construct a simple regression model $\tilde{R}$ using support vector regression (SVR) with Gaussian Kernel. The sparse sampling $m(x=x_k,z)$ and $v$ at its corresponding location $v(x=x_k,z)$ are discretized into $N$ training samples:
\begin{equation}
\left(\left(
\begin{array}{cc}
v^{(i)}  \\
     z^{(i)} 
\end{array}
\right),m^{(i)}\right),
\label{eq:svr}
\end{equation}
where $i=1,2,..., N$.
The model $\tilde{R}_{\theta}(\cdot)$ is trained by minimizing
\begin{equation}
\sum_{N}\left\{\tilde{R}_{\theta}(v^{(i)},z^{(i)})-m^{(i)}\right\},
\label{eq:svrloss}
\end{equation}
and then applied on $v(x,z)$ to build the dense reconstruction $\tilde{R}_{\theta}(v(x,z),z)$. The dense reconstruction provides the global information of $m$, but it is inaccurate due to the simplification of the rock-physics model. To account for the inaccuracy, we add the well log data $m(x=x_k,z)$ as the sparse sampling. It is combined with the dense reconstruction to composite the pseudo label:

\begin{equation}
\tilde{m}(x,z)=\lambda_1\overbrace{\tilde{R}_{\theta}(v(x,z),z)}^{dense{\ } reconstruction}+\lambda_2\overbrace{\sum_{i}m(x,z)*\delta({x_{k})}}^{sparse{\ }sampling}
\label{eq:pseudo}
\end{equation}
where $\lambda_1$ and $\lambda_2$ are the weight for dense reconstruction and sparse sampling. The dense reconstruction provides global but inaccurate information while the sampling provides accurate but local information.

 The end-to-end network takes the geophysical measurements, such as seismic and EM data, as the inputs and generate geophysical properties $m_{pred}$. With the pseudo labels $\tilde{m}$, the network $g$ can be trained with the loss function $\mathcal{L}(m_{pred}, \tilde{m})$ to approximate the inverse mapping $f^{-1}(\cdot)$.
\begin{figure}[ht]
\centering
\includegraphics[width=0.95\columnwidth]{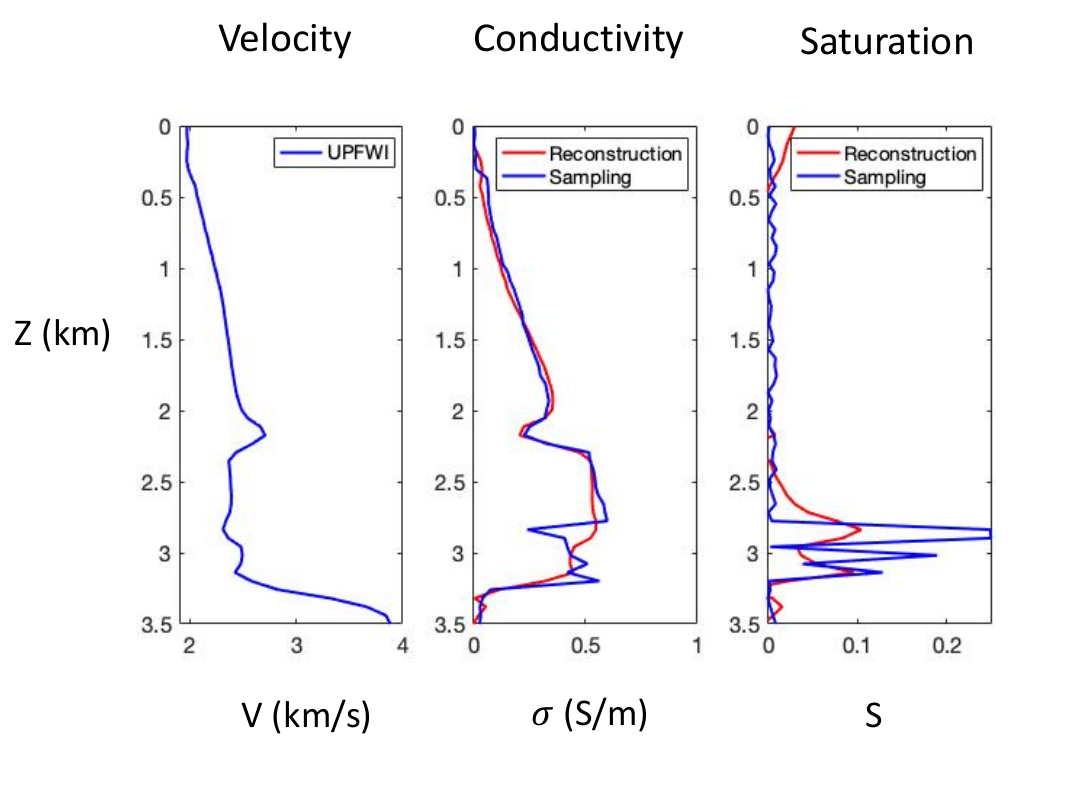}
\caption{ Examples of profiles: the velocity profile provided by UPFWI, the sparse samplings provided by well logs and the dense reconstructions provided by SVR of conductivity and $\mathrm{CO_2}$ saturation models.}
\label{fig:svr_well.pdf}
\end{figure}


\section{Numerical Tests}
In this section, we apply this method to the Kimberlina reservoir dataset. The original geophysical properties were developed under DOE’s National Risk Assessment Program (NRAP) based on a potential $\mathrm{CO_{2}}$ storage site in the Southern San Joaquin Basin of California~\citep{Development-2021-Alumbaugh}. In this data, there are 780 samples and each sample contains a set of seismic and EM data as geophysical measurements, velocity, conductivity, and $\mathrm{CO_{2}}$ saturation models as properties and two well log data that provides CO$_{2}$ saturation and conductivity. In our experiments, 750 samples are set as training set and the rest are the validation set.

\subsection{Kimberlina Data}
The saturation, conductivity, and velocity models are with the size of $59\times100$. The grid is 60 $m$ in all dimensions. Two well logs are located at 2 $km$ and 4 $km$. There are 5 seismic sources placed evenly on the 2D spatial grid over the surface with a shot interval of 1.2 $m$. Seismic data are simulated using the finite-difference method~\citep{moczo2007finite}. Each of them captures vibration signals as time-series data of length 1,001 with a time spacing of 0.005 $s$. EM data are simulated by finite-difference method~\citep{commer2008new} with two sources location at $x=2.5$ $km$, $z=3.025$ $km$ and $x=4.5$ $km$, $z=2.5$ $km$. There are 8 source frequencies from 0.1 to 8.0 Hz and the data with each frequency has a real part and an imaginary part. Both the seismic and EM data are collected by 100 receivers uniformly distributed over the 2D earth surface with a receiver interval of 60 $m$. 

.

\subsection{Workflow}
\textbf{Stage 1:} The velocity models are provided by UPFWI as in Fig.~\ref{fig:SVR_data.pdf}a. The resolution of the UPFWI velocity models is lower than the true velocity models due to the limitation of the frequency in full waveform inversion~\citep{schuster2017seismic}. 

\textbf{Stage 2:} We use two well logs at $x=2$ $km$ and $x=4$ $km$ as the sparse samplings ($\frac{1}{50}$$\times$full$\ $labels) and the UPFWI velocity models (see Fig.~\ref{fig:svr_well.pdf}) at the corresponding location in the training of SVR to predict the dense reconstructions. The predicted dense reconstructions are shown in Figs.~\ref{fig:SVR_data.pdf}b and~\ref{fig:SVR_data.pdf}c and their vertical profiles are shown in Fig.~\ref{fig:svr_well.pdf}. We can see the dense reconstruction are inaccurate, especially the reservoir area in the conductivity model (see Green boxes in Fig.~\ref{fig:SVR_data.pdf}b) and the high saturation area in $\mathrm{CO_2}$ saturation model (see Yellow boxes in Fig.~\ref{fig:SVR_data.pdf}c). Then we combine the inaccurate dense reconstructions and the accurate sparse samplings to construct pseudo labels with Eq.~(\ref{eq:pseudo}). The pseudo labels are fed into an end-to-end network $g$ to learn the mapping from the seismic and EM data to conductivity and CO$_2$ saturation.




\subsection{Main Results}
\begin{figure*}[ht]
\centering
\includegraphics[width=2\columnwidth]{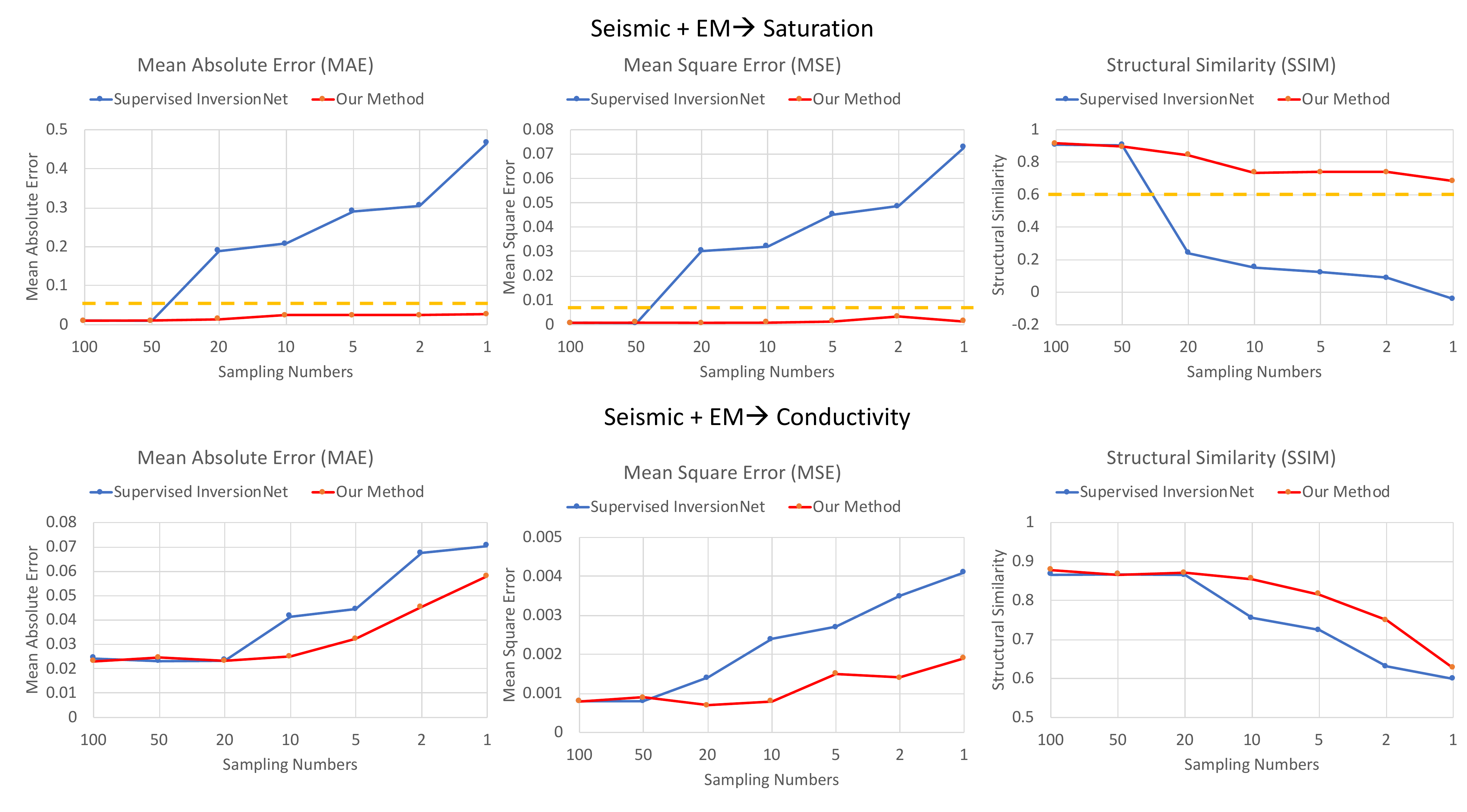}
\caption{\textbf{Weakly Supervised Multiple Geophysics Inversion (ours) vs. Supervised InversionNet~\citep{wu2019inversionnet}}. Our method achieves better performance e.g. lower Mean Absolute Error (MAE) and higher Structural Similarity (SSIM).}.
\label{fig:co2chart.pdf}
\end{figure*}

\begin{figure*}[ht]
\centering
\includegraphics[width=2\columnwidth]{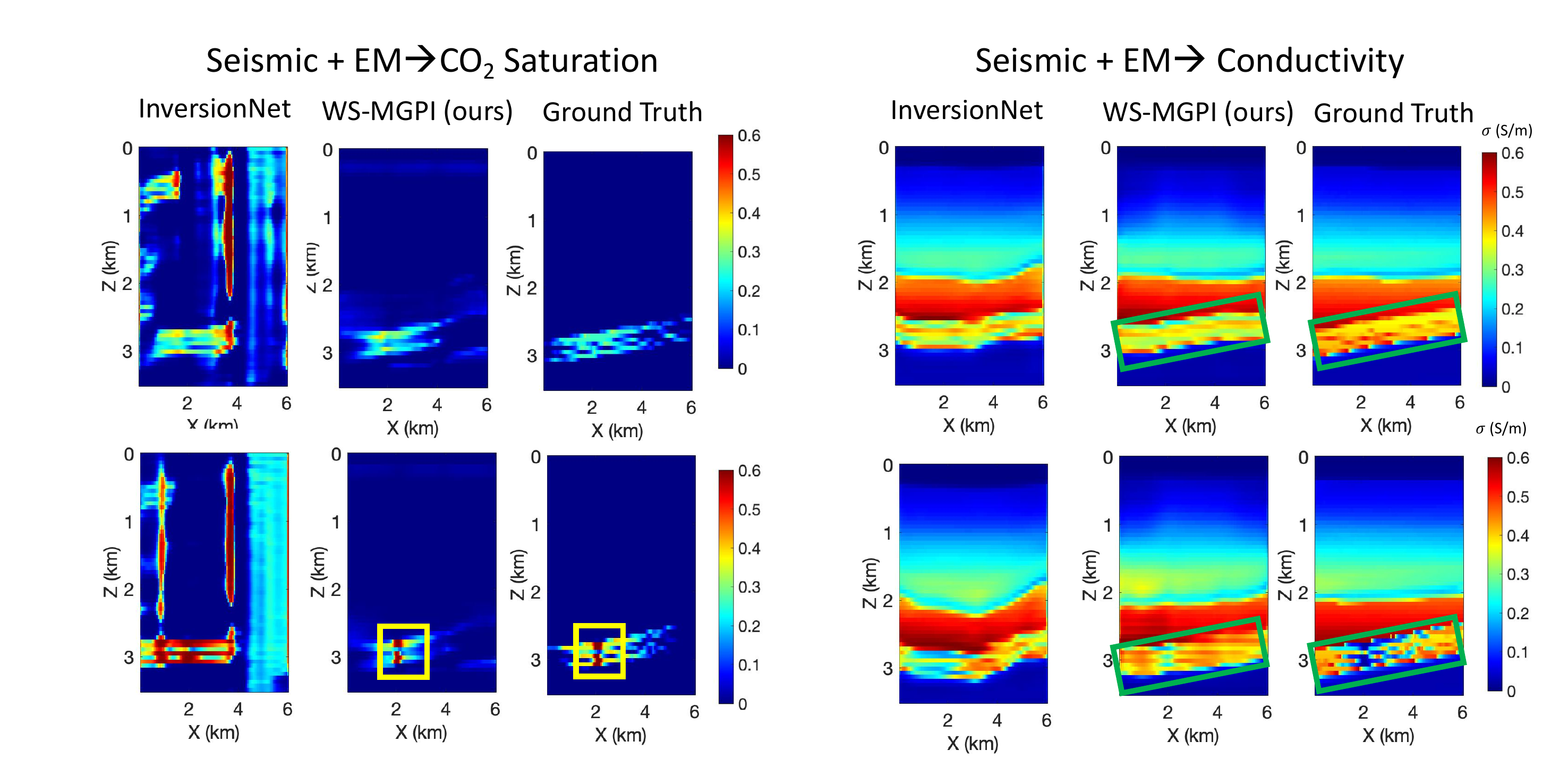}
\caption{Comparison of InversionNet and WS-MGI (ours) on inverted $\mathrm{CO_2}$ saturation and conductivity models when sampling number equal to 2.}
\label{fig:Main_result.pdf}
\end{figure*}

The mean-square errors~(MSE), mean-absolute errors~(MAE), and structural similarity~(SSIM) are used for evaluating the conductivity and saturation. We compare our methods with the supervised InversionNet method~\citep{wu2019inversionnet,zeng2021inversionnet3d}. There are 100 samples for the full label, we gradually decrease the number of samples and evaluate the performance of the methods when the sampling becomes more and more sparse. Fig.~\ref{fig:co2chart.pdf} compares the results with the supervised InversionNet and our method on two scenarios:

\textbf{Seismic+EM${\rightarrow}$$\mathbf{CO_2}$ Saturation}: 
In this scenario, seismic and EM data are set as the input measurement and the target property $m$ is {CO$_2$} saturation. The ratio between the weight $\lambda_1$ and $\lambda_2$ is set as 1. When the sampling number is less than 20 ($\frac{1}{5}$$\times$full$\ $labels), the performance of the InversionNet quickly degrades, the MAE becomes higher than 0.2, MSE increase to 0.03 and SSIM decrease to 0.2. But our method always keeps MAE less than 0.05, MSE less than 0.01, and SSIM higher than 0.6. Examples of the results are shown in Fig.~\ref{fig:Main_result.pdf}. The saturation models given by InversionNet contain large amount of artifacts in the background. The results with our WS-MGI method are consistent with the ground truth. Moreover, the high saturation zone in the yellow box is inverted clearly. 

\textbf{Seismic+EM${\rightarrow}$Conductivity}: In this scenario, seismic and EM data are the input measurement, and the target property $m$ is conductivity. The ratio between the weight $\lambda_1$ and $\lambda_2$ is set as 10. The relationship between the conductivity and EM data is governed by the PDE. When the sampling number decreases, the performance of InversionNet decreases slower than the saturation. However, our methods still have lower MAE, lower MSE, and higher SSIM than those of the InversionNet for all the sampling numbers. In Fig.~\ref{fig:Main_result.pdf}, the thin layers in the blue boxes are reconstructed much better in our result than the one obtained using InversionNet.

\section{Conclusions}
In this paper, we proposed Weakly Supervised Multiple Geophysics Inversion (WS-MGI) that solve multi-physics inversion problem with sparse samplings. With pseudo labels built from the sparse labeling of the properties, we can train an end-to-end network that learns the mapping from the measurement to the property. This network enables the inversion of geophysical properties that only have an implicit relationship with the measurement. Moreover, solving the multi-physics inversion in a weakly supervised way saves the extremely high cost of the label collection, which is much more practical than the previously existing supervised inversion methods.


\section{ACKNOWLEDGMENTS}

This work was funded by the Los Alamos National Laboratory~(LANL) - Laboratory Directed Research and Development program under project number 20210542MFR and by the U.S. Department of Energy~(DOE) Office of Fossil Energy’s Carbon Storage Research Program via the Science-Informed Machine Learning to Accelerate Real Time Decision Making for Carbon Storage~(SMART-CS) Initiative.

\twocolumn
\onecolumn

\bibliographystyle{seg}  
\bibliography{example}

\end{document}